\documentclass[12pt]{article}
\usepackage{graphicx}
\usepackage{bm}
\usepackage{latexsym}
\newcommand{\beq}{\begin{equation}}
\newcommand{\eeq}{\end{equation}}
\newcommand{\bc}{\begin{center}}
\newcommand{\ec}{\end{center}}
\newcommand{\eeqa}{\end{eqnarray}}
\newcommand{\beqa}{\begin{eqnarray}}

\newcommand{\ra}{\rightarrow}
\newcommand{\na}{\nabla}

\newcommand{\al}{\alpha}
\newcommand{\be}{\beta}
\newcommand{\ga}{\gamma}

\newcommand{\de}{\delta}

\newcommand{\ep}{\epsilon}

\newcommand{\si}{\sigma}

\newcommand{\ta}{\tau}

\newcommand{\ph}{\phi}

\newcommand{\om}{\omega}
\newcommand{\ed}{\end{document} }
\begin{document}
\baselineskip=20pt
\raggedright
\parindent3em
\title{Comparison of spin effects and self forces on the electron}

\author{Richard T. Hammond\thanks{rhammond@email.unc.edu }\\
Department of Physics\\
University of North Carolina at Chapel Hill\\
}
\date{\today}
\maketitle
\begin{abstract}
{The relativistic equation of motion of a particle with spin is derived. The effects of the spin are compared to the effects of the self force. For low energies the spin effects are shown to be two orders of magnitude larger than the self force. For higher energies is is shown that the self forces are bigger, and the overall magnitude of the effects are compared.}

\end{abstract}

\section{Introduction}

With reports of laser intensities of $10^{22}$ W cm$^{-2}$, and expectations of an increase by at least an order of magnitude,\cite{michigan}  the classical equation of motion of electrons in electromagnetic fields is gaining renewed interest. When considering the motion of electrons in such extreme fields, these particles quickly become free and attain energies that vastly exceed their rest energies. The correspondence principle, with such extremely high principle quantum numbers, clearly indicates the validity of the classical approximation, where not only radiation reaction effects become important,\cite{hammond06} but the effect of the interaction of the intrinsic magnetic moment of the particle and the external electromagnetic field becomes large.\cite{walser} 

An unsolved problem relates to the comparison of the intrinsic spin effects to the self self forces. Specifically, we need to understand at what energies spin effects dominate and if, and when, self forces predominate. A simple calculation in the low velocity limit indicates that the spin effects are important. Let us consider the ratio, $R_o$, self force to the spin force in the low velocity limit,

\beq
R_o=\left|{2e^2 \dot a/3\over {\bm \na}({\bm\mu }\cdot{\bm B })}\right|
,\eeq
and use $\dot a= e\dot E/m$, where the dot indicates the time derivative, $e$ is the charge, $a$ is the acceleration, $\mu$ is the magnetic dipole moment, and $c=1$. For an electromagnetic wave of the form $E\cos(kx-\om t)$ this yields

\beq\label{Ro}
R_o=\al  
\eeq
where $\al$ is the fine structure constant. This result shows that the spin force is two orders of magnitude larger than the radiation reaction force. Of course this result holds in the low velocity limit, and indicates that the spin force should be examined more carefully. If one naively (and incorrectly) replaces $da/dt$ by $da/d\ta$ in the above, than the quotient in (\ref{Ro}) is multiplied by $v^0$, which becomes large for high energies. This indicates that at high energies the self forces may become larger than the spin forces, and a more careful analysis in the following will show that this is the case.

The point, and an objective of this paper, is to find the relativistic generalization of this result. This will be achieved by first developing the classical equation of motion with spin, ignoring radiation reaction effects. Then the equation of motion with self forces will be examined, ignoring spin, and the results will be compared. In this approximation  we ignore the radiation reaction effects of the spin part, which is a small part of a small part.

An early detailed work on this subject is given by Bjabha and Corben.\cite{bhabha} This subject was also treated in the book of Barut,\cite{barut} who returned to the problem of radiation reaction years later.\cite{barut89} In the 1960s, a classical equation of motion with a spin interaction was derived by starting with the quantum mechanical equation, and obtaining the classical limit by disregarding the commutators.\cite{good}
A year later, following an attempt by Pauli, an asymptotic solution to the Dirac equation was developed using the WKB method for a particle with an anomalous moment,\cite{rubinow} and an anomalous moment was used to explore classical electrodynamics resulting from a nonlinear Dirac equation.\cite{ranada} A comprehensive review of particle-laser interaction in high intensities has been published recently.\cite{salamin}

This article derives the equation of motion from an action principle, but also derives the field equations that follow from the same action. The generalized BMT equation is also derived. A few definitions will help clarify the discussion below. The spin tensor is defined in terms of the spin vector $S^\al$ according to

\beq
S_{\mu\nu}=\ep_{\mu\nu\al\be}S^\al v^\be
\eeq
where $\ep_{\mu\nu\al\be}$ is the totally antisymmetric tensor. Equivalently 

\beq
S_\mu=\frac12S^{\al\be}v^\si\ep_{\si\al\be\mu}
.\eeq
Other useful definitions are the magnetic moment tensor, 

\beq\label{mu}
\mu^\al=\frac emS^\al
\eeq
and the electromagnetic dual $^*\!\!F^{\mu\nu}= \frac 12 F_{\al\be}\ep^{\al\be\mu\nu}$.

Finally, the traditional equation of motion of a charged particle in an electromagnetic field is given by

\beq\label{eqmno}
{dp^\mu\over d\tau}=eF^{\mu\si}v_\si
\eeq
where $p^\mu=mv^\mu$. 

\section{Equation of motion with spin}

We know that the force on a magnetic dipole in a magnetic field is given by

\beq\label{magforce} 
{\bm f}={\bm \na}\left({\bm \mu}\cdot{\bm  B}\right)
.\eeq
This well known result is derived by integrating the Biot-Savart law around a small current loop and defining the magnetic moment as a line integral in the usual way. It is not a valid definition for an elementary particle, since we know that the magnetic moment cannot be explained by a small loop current. Therefore we introduce the interaction directly into the action. The relativistic generalization of the scalar ${\bm \mu}\cdot{\bm  B}$ is

\beq\label{FS}
{e\over2m}F_{\mu\nu}S^{\mu\nu}
,\eeq
which implies the generalization of (\ref{magforce}) is $f^\mu=(e/2m)(F_{\mu\nu}S^{\mu\nu})^{,\mu}$ and therefore the naive relativistic generalization of (\ref{eqmno}) is $dp^\mu/ d\tau=eF^{\mu\si}v_\si +f^\mu $. This is incorrect because it fails to obey the condition that the velocity is orthogonal to the acceleration\footnote{The identity $v_\si v^\si=1$ implies that 
$v_\si dv^\si/d\ta=0$.} One may correct this by hand, by generalizing $f^\mu$ to
\beq\label{fmu}
f^\mu={e\over2m}[(F_{\al\be}S^{\al\be})^{,\mu}-(F_{\al\be}S^{\al\be}v^\mu)_{,\ph}v^\ph]
.\eeq

This result will be verified by starting from the action

\beq
I= -m\int ds-e\int A_\si dx^\si -{1\over16\pi}\int d^4xF_{\mu\nu}F^{\mu\nu}
-{e\over 2m}\int ds F_{\mu\nu}S^{\mu\nu}
\eeq
where the particle moves along the trajectory parametrized by $s$. This is the usual action of electromagnetism with the additional spin interaction added. The equation of motion may be obtained by considering variations with respect to the path. Using 

\beq\label{desmunu}
\de S^{\mu\nu}=S^{\mu\nu}_{\ \ ,\si}\de x^\si
\eeq
we have\footnote{
An important choice has been made here. One may 
use $F_{\mu\nu}S^{\mu\nu}=2^*\!\!F_{\mu\nu}S^\mu v^\nu$ and then consider 
$\de S^\mu=S^\mu_{ ,\si}\de x^\si$. In the low velocity limit this predicts that the spin force is given by ${\bm f}= -{d\over dt}({\bm E}\times{\bm \mu})/c$, which is the incorrect limit. The correct limit is given by (\ref{flim}).}

\beq\label{eqmspin}
{dp^\mu\over d\tau}=eF^{\mu\si}v_\si+f^\mu
,\eeq
where
$f^\mu$ is given by (\ref{fmu}).
After the variation, we take $s$ to be the proper time $\tau$. This equation may also be written as

\beq
{d\over d\tau}\left(v^\mu\tilde m\right)
=eF^{\mu\si}v_\si+(^*\!F_{\al\be}S^\al v^\be)^{,\mu}
.\eeq
This form shows that one may consider $\tilde m=m+^*\!\!F_{\al\be}S^\al v^\be$ as an effective mass, which can be seen directly from the action. 

In the low velocity limit one may show that,
calling $f^n\ra{\bm f}$,

\beq\label{flim} 
{\bm f}={\bm \na}\left({\bm \mu}\cdot({\bm  B}-{\bm  v}\times{\bm  E})\right)
.\eeq
This is a necessary result, showing that the action reduces to the correct limiting form. The field equations that follow from this action are discussed in the Appendix.
 
\section{Generalized BMT equation}

The BMT equation starts from \cite{bargmann}

\beq\label{dsdt}
{d S^\al\over d\ta}={ge\over 2m}\left(
F^{\al\be}S_\be+v^\al v_\mu S_\si F^{\si\mu}\right)
-v^\al S_\be {d v^\be\over d\ta}
.\eeq
The conventional BMT equation is obtained by using (\ref{dsdt}) with $f^\mu=0$. To obtain the generalized BMT equation we retain $f^\mu$ and obtain

\beq
{d S^\al\over d\ta}={ge\over 2m}F^{\al\be}S_\be+{ge\over2m}(g-2)v^\al S_\al v_\be F^{\al\be}
-v^\al S_\be f^\be/m
.\eeq

It is helpful to write the equations in nondimensional form. We let $x^\mu\ra x^\mu/L$ and $t\ra tc/L$ . For example, if we consider a plane wave, ${\bm E} =E \cos(kz-\om t){\bm \hat x}\ra E \cos(z-t){\bm \hat x}$ if $L=\lambda/2\pi$. Also, $S\ra S/\hbar$ and $F^{\mu\nu}\ra F^{\mu\nu}/E$. Then the equations of motion for an electron, $(g=2)$, become,

\beq\label{trans}
{dv^\mu\over d\ta}=av_\si F^{\mu\si} +Af^\mu
\eeq
and

\beq\label{spin}
{dS^\mu\over d\ta}=aS_\si F^{\mu\si} -Av^\mu S_\si f^\si
\eeq
where
\beq
f^\mu= (F_{\al\be}S^{\al\be})^{,\mu}-{d\over d\ta}(F_{\al\be}S^{\al\be}v^\mu)
.\eeq

All terms are dimensionless, and the strengths are determined by the dimensionless constants (bringing back $c$),

\beq 
a={eEL\over mc^2}\ \ \ \mbox{and} \ \ A={eE\hbar\over m^2c^3}
.\eeq

The ratio $A/a=\hbar/(mcL)$  shows that the spin effects on the translational equation of motion are generally small but, for an electron mass, become appreciable for hard x-rays and important for gamma rays, and may even be dominate for high energy gamma rays such as observed by EGRET.

\section{Radiation Reaction}

As mentioned in the introduction, intensities of $10^{22}$
W cm$^{-2}$ have been reached, and this number is expected to go even higher. At such extreme conditions, the radiation reaction force is expected to become important, and we will examine the intensities where the onset of the self force becomes important before comparisons with the spin is made.

It is helpful to begin with the relativistic  equation of motion, (\ref{eqmno}), and consider an electromagnetic wave of the form  

\beq\label{gx}
{\bm E} =Eh(z-t)\bm{ \hat x}
\eeq
and

\beq\label{yy}
{\bm B} =Eh(z-t)\bm{ \hat y}.
\eeq
This represents a plane wave of amplitude $E$, polarized in the $x$ direction, described by any dimensionless function $h$. With this (\ref{eqmno}) gives,

\beq\label{v0}
{dv^0\over d\ta}=ahv^1
\eeq

\beq\label{v1orig}
{dv^1\over d\ta}=ah(v^0-v^3)
\eeq

\beq\label{v2}
{dv^2\over d\ta}=0
\eeq

\beq\label{v3}
{dv^3\over d\ta}=ahv^1
.\eeq

We note that (\ref{v0}) and (\ref{v3}) imply,

\beq\label{v03}
v^0=1+v^3
\eeq
which leaves the pair

\beq\label{eq2}
{dv^0\over d\ta}=ahv^1
\eeq

\beq\label{v1}
{dv^1\over d\ta}=ah
.\eeq
These imply, 

\beq\label{v01}
v^0=1+(v^1)^2/2
.\eeq
Using the integral of (\ref{v03}) in the right hand side of (\ref{v1}) yields

\beq\label{v1g}
v^1=a\int h(-\tau) d\ta
,\eeq
a fascinating result. It states that the $x$ component of the four velocity is essentially equal to the nonrelativistic three velocity evaluated at the proper time $\ta$. With this,
(\ref{v03}), and (\ref{v01}), the relativistic solution is completely determined in terms of the nonrelativistic solution for an arbitrary 1D wave form $h(z-t)$.
Having an analytical expression for the velocity is useful for looking at both the self force and the spin force in an iterative approach. 

The Lorentz Abraham Dirac equation of motion, which includes the self force, is
\footnote{We assume $v_\si v^\si=1$. In the literature, some take $v_\si v^\si=-1$, which changes signs in the self force.}

\begin{equation}\label{lad1}
{dv^\sigma\over d\tau}=aF^{\si\mu}v_\mu+b(\ddot v^\si+\dot v^\nu\dot v_\nu v^\si)
\end{equation}
where  the dimensionless parameter $b\equiv c\ta_0/L$, $\ta_0=2e^2/(3mc^3)$ and the overdots imply differentiation with respect to the proper time.  The LAD equation fell under bad times due to the the runaway solutions, pre-acceleration issues, or the apparent need to invoke non-zero size particles.\cite{ori}  For a nice introduction to the issues, with historical notes, one may consult Rohrlich\cite{rohrlich}, which has many of the seminal references and discusses the distinction between ``self forces'' and ``radiation reaction forces.'' More recent work considers the problem in various dimensions,\cite{galtsov}\cite{kazinski} the effect of the magnetic dipole,\cite{meter}\cite{heras}, connections to QED\cite{higuchi} and vacuum polarization,\cite{binder}, mass conversion\cite{bosanac}, and hydrogenic orbits.\cite{cole}

\begin{figure}[!h]
\includegraphics[width=0.9\textwidth]{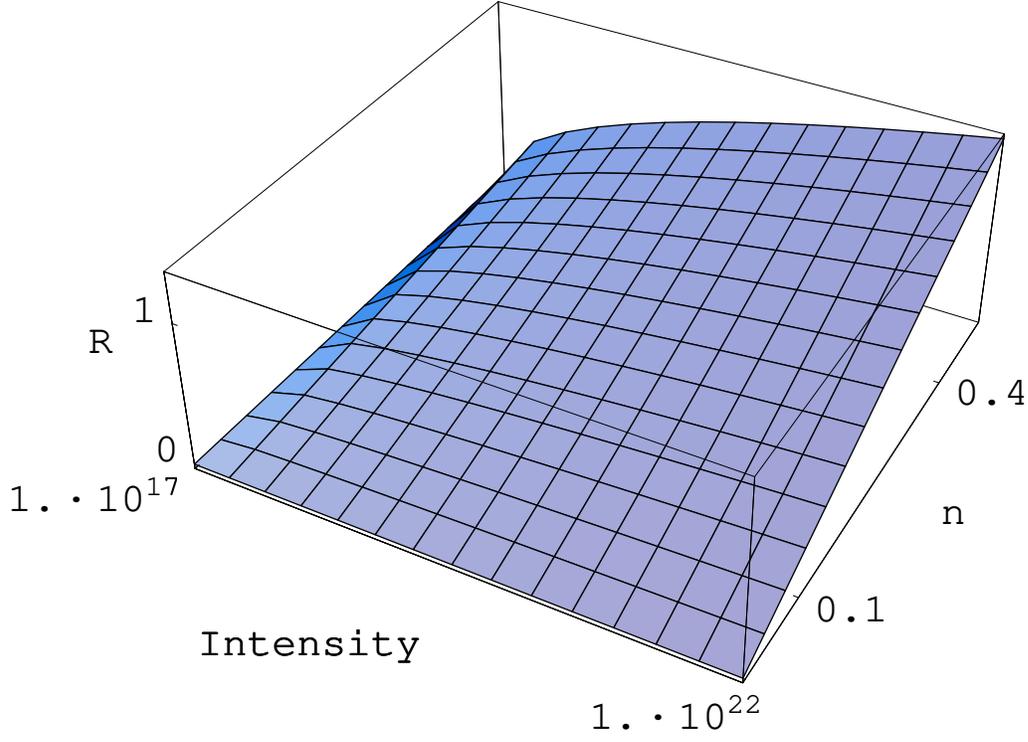}
\caption{The ratio $R$ as a function of wavelength and intensity, with $n=\lambda/L$.}
\label{R3D}
\end{figure}

The Landau Lifschitz Rohrlich equation is  obtained
by starting with (\ref{lad1}) but using 

\beq
{dv^\si\over d\tau}=aF^{\si\mu}v_\mu
\eeq
in the terms multiplied by $b$, which leads to,

\beq\label{LLR}
{dv^\si\over d\tau}= aF^{\si\mu}v_\mu+ br^\si
,\eeq
where

\beq\label{r}
r^\si\equiv  a\dot F^{\si\mu}v_\mu
+a^2(F^{\si\ga}F_\ga^{\ \ph}v_\ph+F^{\nu\ga}v_\ga F_\nu^{\ \ph}v_\ph v^\si)
.\eeq
We may now compare this equation to (\ref{trans}). Since $a>>b$, we may look at the ratio of  self over spin force (ratio of constants only)

\beq\label{R}
R={ba^2\over A }=\frac43\al a
\eeq
which shows for large fields, for any given wavelength, eventually the radiation reaction dominates. In fact this ratio becomes unity at $I\approx 10^{21}$W cm$^{-2}$ for $L=5\times 10^{-5}$. Figure \ref{R3D} shows $R$ on the vertical axis as a function of intensity in watts cm$^{-2}$ and $n\equiv\lambda/L$. This also shows that, for a fixed intensity, the spin effects become important for short wavelengths.

\section{Solutions}

\subsection{Self force}

It is helpful to begin by examining the effect of any force, ${\cal F}^\si$, 
that is multiplied by a small parameter $\Lambda$,

\beq\label{rel}
{dv^\si\over d\ta}=aF^{\si\nu}v_\nu+ \Lambda{\cal F}^\si
\eeq
 and  to expand the solution in terms of $\Lambda$,

\beq
v^\si={_0v}^\si+\Lambda({_1v}^\si)+...
\eeq
where ${_0v}^\si$ is the solution with ${\cal F}^\si=0$, ${_1v}^\si$ is ${\cal O}(\Lambda)$, and so on. Then, for the plane polarized field used above, the zero order solutions are given by (\ref{v03}), (\ref{eq2}), and (\ref{v1}), and the ${\cal O}(b)$ equations are,

\beq\label{v0r}
{d{_1v}^0\over d\ta}=ah(_1v^1)+{\cal F}^0({_0v})+{\cal O}(b^2)
\eeq

\beq\label{v1rr}
{d{_1v}^1\over d\ta}=ah((_1v^0)-{_1v}^3)+{\cal F}^1({_0v})+{\cal O}(b^2)
\eeq

\beq\label{v3rr}
{d{_1v}^3\over d\ta}=ah(_1v^1)+{\cal F}^3({_0v})+{\cal O}(b^2)
.\eeq

 Now let us suppose $\Lambda{\cal F}^\si =br^\si$, which is defined by (\ref{r}). With this, it can be seen that $_1v^0-_1v^3=1 +{\cal O}(b^2)$. Calling $\ph\equiv {_0}v^1$ and dropping the subscripts so that $_1v^\si\ra v^\si$  we  have,

\beq\label{v0ob}
{dv^0\over d\ta}=ahv^1+a\dot h\ph -a^2h^2\ph^2/2+{ \cal O}(b^2)
\eeq
and

\beq\label{v1ob}
{dv^1\over d\ta}=ah+a\dot h -a^2 h^2\ph+{ \cal O}(b^2)
\eeq

\beq
{dv^3\over d\ta}=ahv^1+a\dot h\ph +a^2h^2(1-\ph^2/2)+{ \cal O}(b^2)
\eeq

\begin{figure}
\begin{minipage}[t]{8cm}
\includegraphics[width=0.9\textwidth]{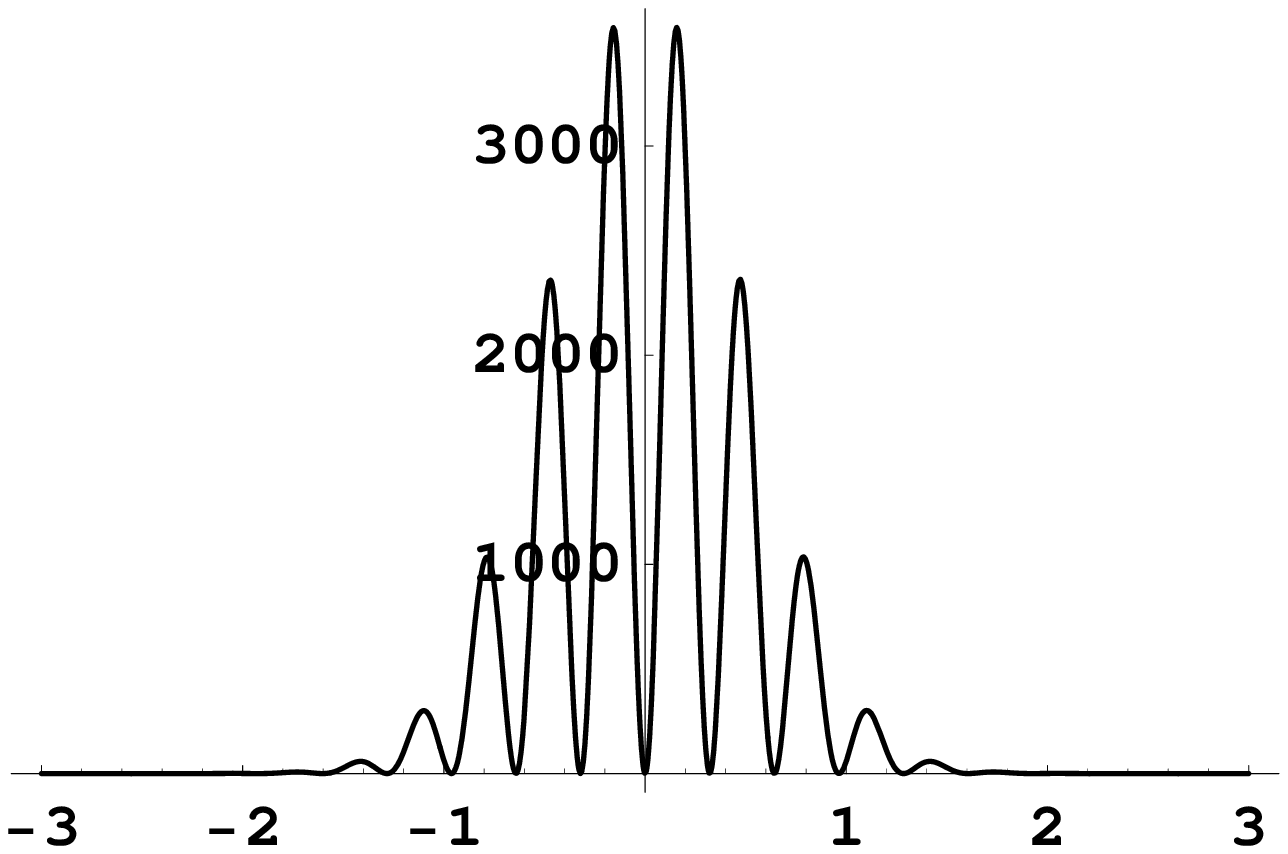}
\caption{$v_z$ plotted against the proper time for $I=10^{23}$W cm$^{-2}$.}
\label{proper}
\end{minipage} \hfill
\begin{minipage}[t]{8cm}
\includegraphics[width=0.9\textwidth]{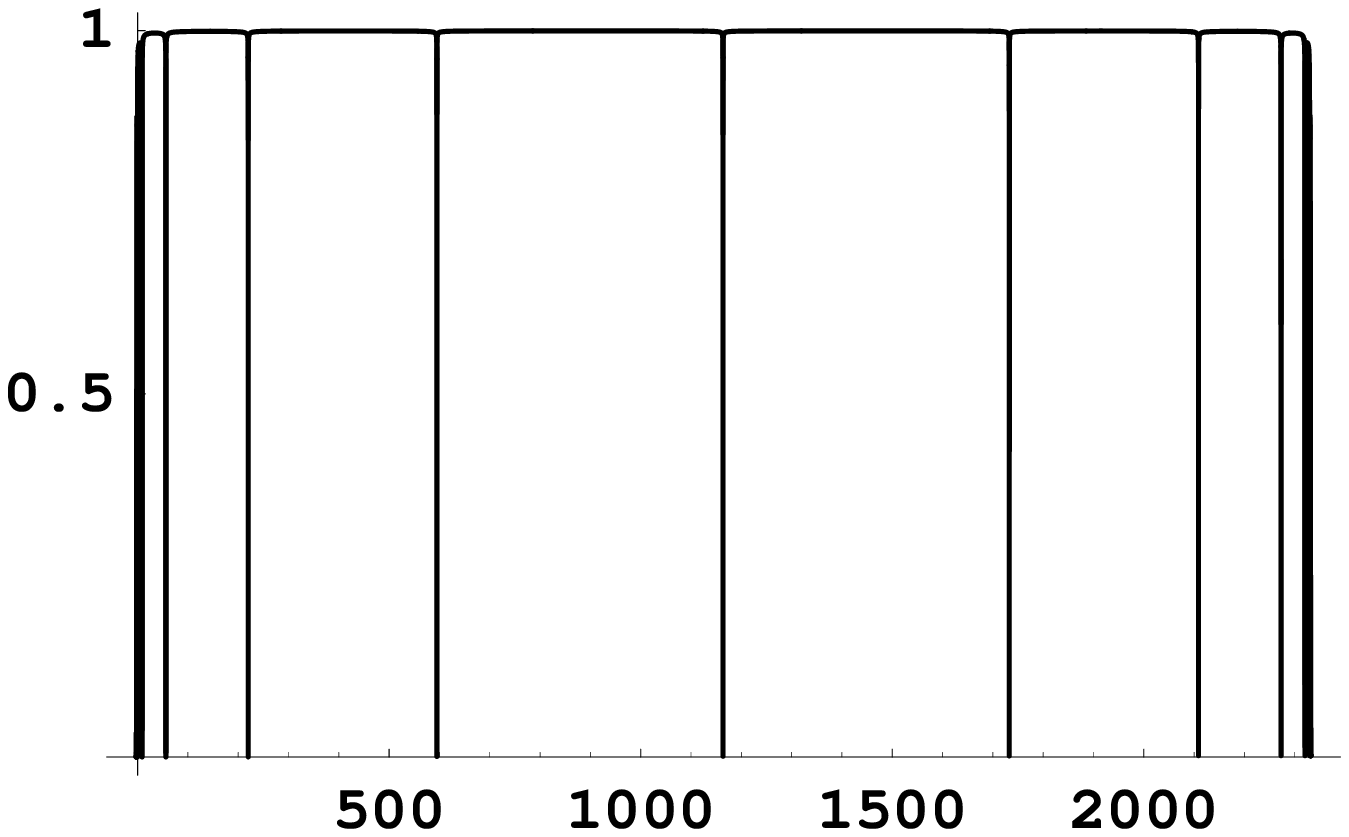}
\caption{$dz/dt$ vs. $t$, corresponding to Fig. \ref{proper}.}
\label{vvst}
\end{minipage}
\hfill
\end{figure}

\begin{figure}
\begin{minipage}[t]{8cm}
\includegraphics[width=0.9\textwidth]{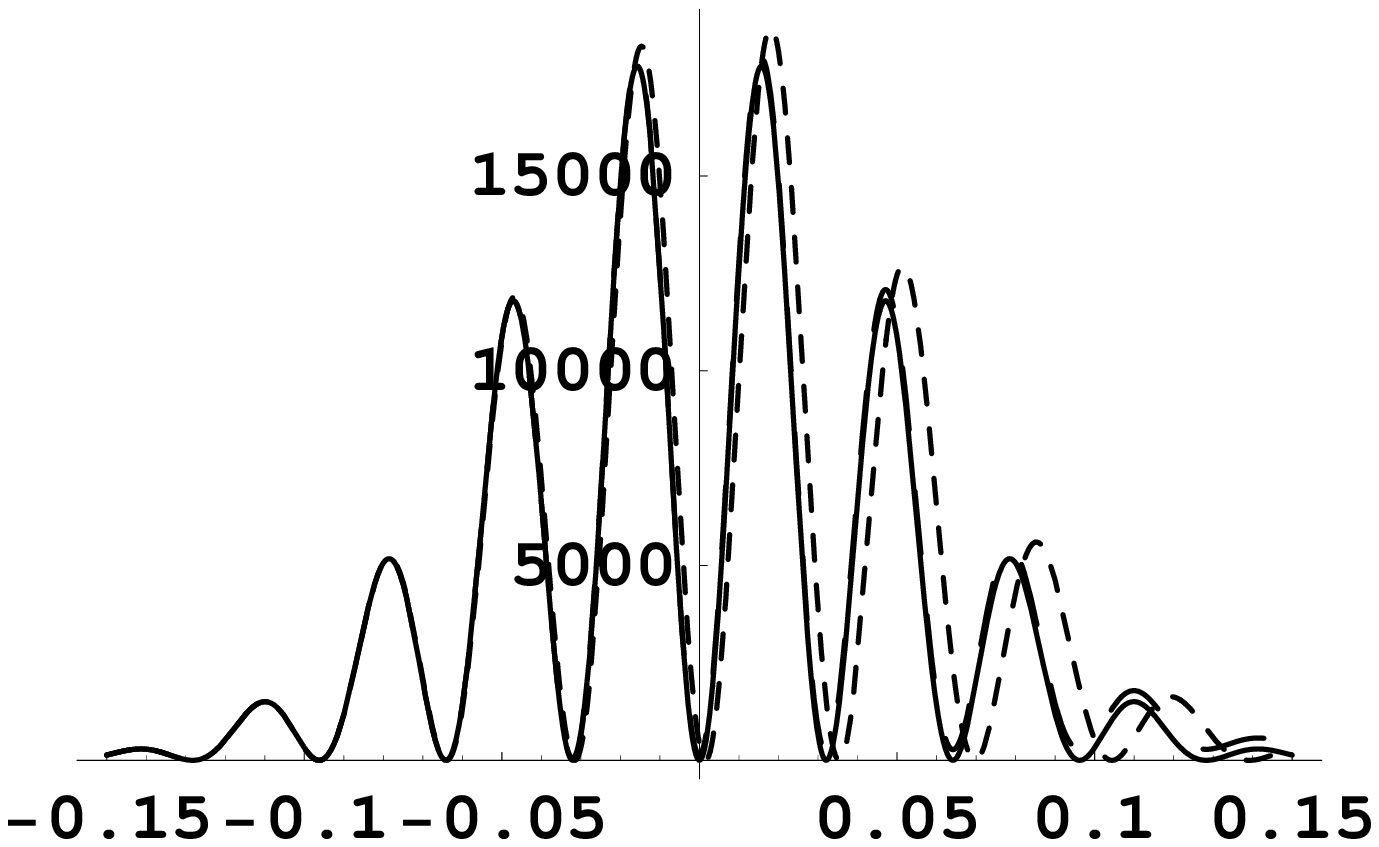}
\caption{$v_0$ plotted against $10t$ with no radiation reaction (solid line), and as a solution to ${\cal O}(b)$ (dotted line).}
\label{g3vs0cos}
\end{minipage} 
\hfill
\begin{minipage}[t]{8cm}
\includegraphics[width=0.9\textwidth]{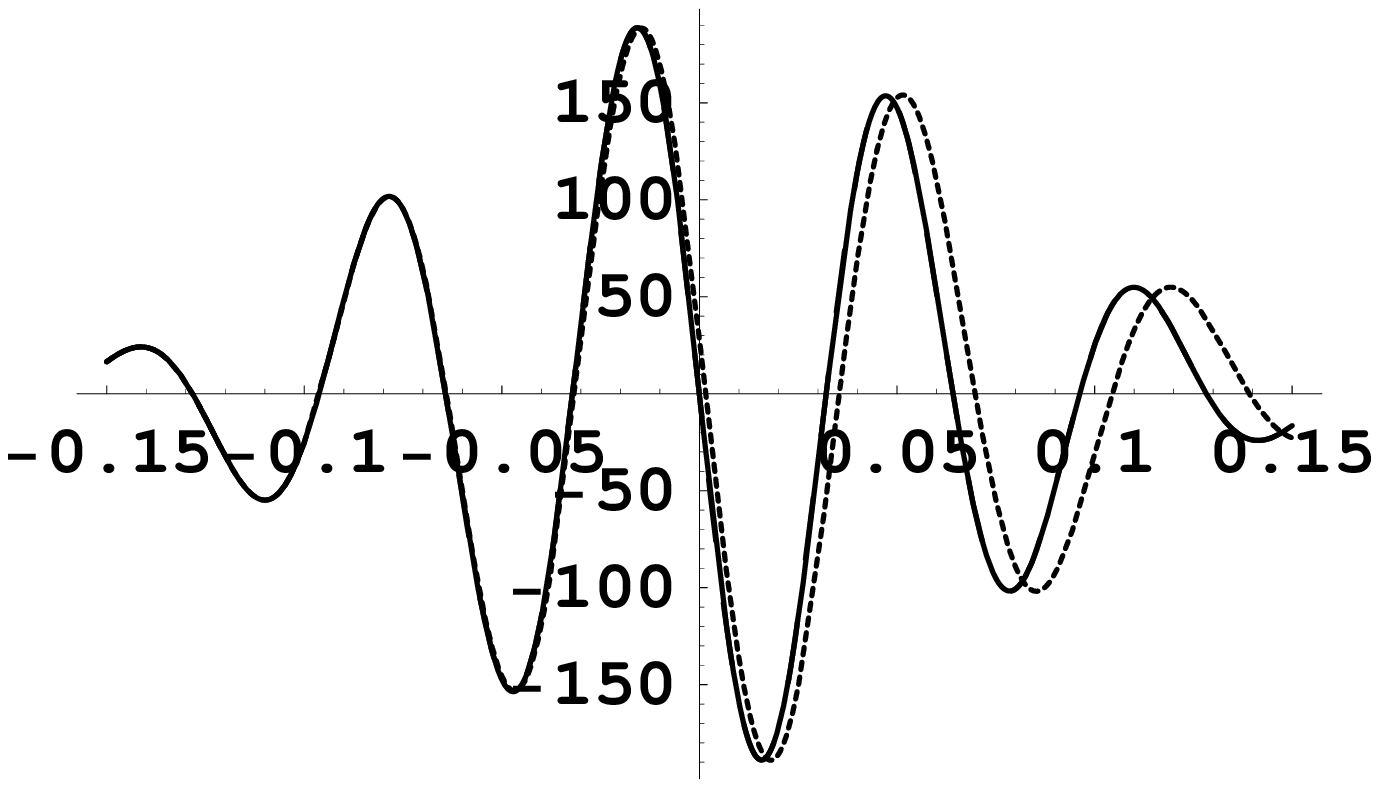}
\caption{$v_1$}
\label{g3vs1cos}
\end{minipage}
\hfill
\end{figure}

As an example, let us consider a pulse of soft x-rays, where we take

\beq
h={e^{-((z-t)/w)^2}\over w}\cos(\Omega(z-t))
\eeq
where the dimensionless $\Omega$ determines the frequency. 
To zero order in $b$ we have,

\beq
{_0v}^1=
\frac{1}{4} a e^{-\frac{1}{4} w^2 \Omega ^2} \sqrt{\pi }
   \left(2+\mbox{erf}\left(\frac{\ta}{w}-\frac{i w \Omega
   }{2}\right)+\mbox{erf}\left(\frac{\ta}{w}+\frac{i w \Omega
   }{2}\right)\right)
.\eeq
 Before considering x-rays it is interesting to consider the case that $\Omega=1=w$ at $10^{23}$W cm$^{-2}$. This is just below the radiation reaction ``threshold,'' but it is interesting to see how strongly relativistic the solution is. This is evident in Figs. \ref{proper} and \ref{vvst}, which show the four velocity as a function of proper time and the corresponding velocity ($dz/dt$) versus $t$ ($t$ is (roughly) laser periods, one may multiply by $L/c$ to obtain time in seconds).
For soft x-rays
$\Omega\sim 100$ (50$\AA$ radiation) and we take $w=1/10$ ($\sim 200$as pulse).
The results are plotted in Figs. \ref{g3vs0cos} and \ref{g3vs1cos} for  $I=5\times10^{23}$  W cm$^{-2}$.

For the simple case of an infinite plane wave a typical solution is

\beq\label{selfv1}
v^1=-a\sin(z-t)+b\left(-a\sin(z-t)+a\cos(z-t) -{a^3\over3}\cos^3(z-t)\right)
+{\cal O}(b^2)
,\eeq
which will be used to compare the effects to spin.

\subsection{Spin}

Consider first the case that $A/a<<1$, valid for all but very short wavelength radiation. The formalism of the previous section may be used and we find in lowest order,

\beq\label{trans0}
{dv^\mu\over d\ta}=av_\si F^{\mu\si}
\eeq
and

\beq\label{spin0}
{dS^\mu\over d\ta}=aS_\si F^{\mu\si}
,\eeq
for the same field that was used previously. We already found (\ref{v03}) -- (\ref{eq2}),
and (\ref{spin0}) shows that
$S^0-S^3=$ const $\equiv s$ and
\
\beq
S^1=as\sin(z-t)
\eeq
and
\beq
S^3={a^2s\over2}\sin^2(z-t)
,\eeq
and $S^2=S^2_{\mbox{in}}$, the incident value of the $y$ component.
This result may be used to evaluate $f^\mu$ in an iterative manner. Using
the above results for the spin  we find, for example,

\beq\label{spinv1}
v^1 =-a\sin(z-t) -A\left(a\sin(z-t)+ AS^2_{\mbox{in}}a\sin2(z-t)\right)
.\eeq
These solutions correspond to somewhat specific incident fields, but the main purpose is to examine the relative weight of the effects of radiation reaction spin forces.
Comparing  (\ref{spinv1}) to (\ref{selfv1}) explicitly verifies the result (\ref{R}).

In summary, the classical equations of motion for a particle with an intrinsic magnetic moment in an electromagnetic field have been derived. They were derived from an action principle in Minkowski spacetime along with the field equations.  The generalized BMT equation was given, and the non-dimensionalized form of the coupled equations was given, a simple approximate solution was exhibited. The effects of the spin force and the self force were considered for high intensity radiation, and conditions where the radiation forces become larger than spin effects were examined.

The results derived here are not readily amenable to be tested in current laboratory experiments, but they tell us when spin is important, the onset of radiation reaction, and the energies and wavelengths for which self forces begin to dominate over spin. These results will be used to guide future work that will calculate observable characteristics, such as radiation emission spectra, of ultrarelativistic particle motion. Another interesting area with this formalism can be applied is the motion of neutral particles with spin in strong magnetic fields, such as the behavior of neutrons near magnetars.

\section{Appendix}

The action has been generalized to describe the interaction of the electromagnetic field with a point dipole, but we should bear in mind that the field equations also follow from the action. In this case we hold the trajectory fixed while considering variation in $A^\mu$. This gives,

\beq
F^{\mu\nu}_{\ \ ,\mu}=4\pi(j^\nu+i^\nu)
\eeq
where we  find the usual 

\beq\label{j}
j^\nu=e\int \de^4(x-x(\ta)) v^\nu d\ta
\eeq
and, using $\de\equiv\de^4(x-x(\ta))$,

\beq\label{i}
i^\nu=\frac em\int dx_\be(\ep^{\mu\nu\al\be}S_\al\de)_{,\mu}
.\eeq

Thus, the equations of electrodynamics have been altered by the additional source $i^\nu$, and it is important to consider the ramifications of any modification to such a successful theory as electrodynamics. First, it is easy to see that since $F^{\mu\nu}_{\ \ ,\mu,\nu}\equiv0$, $(j^\mu+i^\mu)_{,\mu}=0$. The definitions (\ref{j}) and (\ref{i}) show that the four divergence of each of these vanish. This we have the usual conservation of charge.

To see the effect of the new source term, consider the nonrelativistic limit, and consider the source defined over some small region of space. A mulitpole expansion of the vector potential is

\beq A_n(x)=\frac1x\int J_n(x')d^3x'+{{\bm x}\over x^3}\cdot \int {\bm x'}J_n(x')d^3x'+...
\eeq
where $J_n\equiv j_n+i_n$. Now consider some small volume where the contribution
from $j_n$ is negligible (such as the size of a particle). Related calculations may be found elsewhere,\cite{hammond02} and it may shown that the first term gives no contribution and the second leads to,\cite{mckeon}

\beq\label{A}
{\bm A}={{\bm \mu}\times{\bm r}\over r^3}
.\eeq

This new term actually solves a fundamental flaw in electrodynamics. The electron has a magnetic dipole moment which gives rise to a magnetic dipole field. This field can be measured classically, e. g., be detecting the magnetic field of any substance in which the field arises solely from the intrinsic moment (and not from the orbital angular momentum). As is well known, it is impossible to account for this observed field from $j^\mu$ alone: Any charge distribution confined to particle sizes would have to involve linear speeds far in excess of the speed of light. However, (\ref{A}) shows that the new source term gives precisely what
is observed.\footnote{ 
This does not predict the magnetic moment, though. That was put in by hand in (\ref{FS}),  (\ref{mu}) shows that $g$, the gyromagnetic ratio, was taken to be 2.}

As far as QED is concerned, we still assume that the Lagrangian density is
 
\beq
{\cal L}=\overline\psi(i p^\mu-A^\mu\partial_\mu -m)\psi-\frac14F_{\mu\nu}F^{\mu\nu}
.\eeq
 
This gives 
 
\beq
F^{\mu \nu}_{\ \ \ ,\nu}=e\overline\psi\ga^\mu\psi
\eeq
which shows that, quantum mechanically, the interaction is unaffected by the new spin term, although quantum spin effects might manifest themselves on other ways.\cite{hammond95}

\ed